\documentclass[11pt,letterpaper]{article}
\usepackage[lmargin=1.0in,rmargin=1.0in,bottom=1.0in,top=1.0in,twoside=False]{geometry}

\usepackage{fullpage,amssymb,amsmath}
\usepackage{graphicx}
\usepackage{enumerate}
\usepackage[T1]{fontenc}
\usepackage{xcolor}
\usepackage{amsfonts}
\usepackage{comment}
\usepackage[english]{babel}
\usepackage{libertine}
\usepackage[absolute]{textpos}
\usepackage{enumitem}
\usepackage{mathtools}

\definecolor{MyBlue}{RGB}{50,100,200}
\usepackage[pdftex]{hyperref}
\hypersetup{%
  colorlinks=true,
  linkcolor=MyBlue,
  citecolor=MyBlue,
  urlcolor=MyBlue
}

\usepackage[amsmath,thmmarks,hyperref]{ntheorem}
\usepackage{cleveref}

\crefformat{page}{#2page~#1#3}%
\Crefformat{page}{#2Page~#1#3}%
\crefformat{equation}{#2(#1)#3}%
\Crefformat{equation}{#2(#1)#3}%
\crefformat{figure}{#2Figure~#1#3}%
\Crefformat{figure}{#2Figure~#1#3}%
\crefformat{section}{#2Section~#1#3}
\Crefformat{section}{#2Section~#1#3}
\crefformat{chapter}{#2Chapter~#1#3}
\Crefformat{chapter}{#2Chapter~#1#3}
\crefformat{chapter*}{#2Chapter~#1#3}
\Crefformat{chapter*}{#2Chapter~#1#3}
\crefformat{part}{#2Part~#1#3}
\Crefformat{part}{#2Part~#1#3}
\crefformat{enumi}{#2(#1)#3}
\Crefformat{enumi}{#2(#1)#3}

\theoremnumbering{arabic}
\theoremstyle{plain}
\theoremsymbol{}
\theorembodyfont{\itshape}
\theoremheaderfont{\normalfont\bfseries}
\theoremseparator{.}

\newtheorem{theorem}{Theorem}
\crefformat{theorem}{#2Theorem~#1#3}
\Crefformat{theorem}{#2Theorem~#1#3}

\newcommand{\newtheoremwithcrefformat}[2]{%
  \newtheorem{#1}[theorem]{#2}%
  \crefformat{#1}{##2\MakeUppercase#1~##1##3}%
  \Crefformat{#1}{##2\MakeUppercase#1~##1##3}%
}
\newcommand{\newseptheoremwithcrefformat}[2]{%
  \newtheorem{#1}{#2}%
  \crefformat{#1}{##2\MakeUppercase#1~##1##3}%
  \Crefformat{#1}{##2\MakeUppercase#1~##1##3}%
}

\newtheoremwithcrefformat{lemma}{Lemma}
\newtheoremwithcrefformat{proposition}{Proposition}
\newtheoremwithcrefformat{observation}{Observation}
\newtheoremwithcrefformat{corollary}{Corollary}
\newseptheoremwithcrefformat{claim}{Claim}
\newseptheoremwithcrefformat{definition}{Definition}
\theorembodyfont{\upshape}
\newtheoremwithcrefformat{example}{Example}
\newseptheoremwithcrefformat{remark}{Remark}

\theoremstyle{nonumberplain}
\newseptheoremwithcrefformat{conjecture}{Conjecture}
\theoremheaderfont{\scshape}
\theorembodyfont{\normalfont}
\theoremsymbol{\ensuremath{\square}}
\newtheorem{proof}{Proof}

\theoremsymbol{\ensuremath{\lrcorner}}
\newtheorem{clproof}{Proof}

\def\cqedsymbol{\ifmmode$\lrcorner$\else{\unskip\nobreak\hfil
\penalty50\hskip1em\null\nobreak\hfil$\lrcorner$
\parfillskip=0pt\finalhyphendemerits=0\endgraf}\fi}

\newcommand{\eps}{\varepsilon}
\renewcommand{\div}{\ \mathrm{div}\ }

\newcommand{\N}{\mathbb{N}}
\newcommand{\R}{\mathbb{R}}
\renewcommand{\phi}{\varphi}

\newcommand{\wei}{\omega}
\renewcommand{\leq}{\leqslant}
\renewcommand{\geq}{\geqslant}
\renewcommand{\le}{\leqslant}
\renewcommand{\ge}{\geqslant}

\newcommand{\stimes}{\boxtimes}

\usepackage[mathscr]{eucal}
\newcommand{\sC}{\mathscr{C}}
\newcommand{\Cc}{\sC}
\newcommand{\sS}{\mathscr{S}}

\newcommand{\sU}{\mathscr{U}}

\renewcommand{\emptyset}{\varnothing}

\newcommand{\ceil}[1]{\left\lceil{#1}\right\rceil}

\newcommand{\pare}[1]{\left({#1}\right)}
\newcommand{\card}[1]{\left|{#1}\right|}
\newcommand{\set}[1]{\left\{{#1}\right\}}
\newcommand{\range}[2]{\set{{#1},\dots,{#2}}}

\begin{document}

\title{\textbf{Shorter Labeling Schemes for Planar Graphs}%
  \thanks{The preliminary version of this work~\cite{BGP20} was presented at the $31^{\textrm{st}}$ Symposium on Discrete Algorithms, SODA 2020. The work of Micha\l{} Pilipczuk is a part of project TOTAL
    that have received funding from the European Research Council
    (ERC) under the European Union's Horizon 2020 research and
    innovation programme (grant agreement No.~677651). The work of Marthe Bonamy and Cyril Gavoille is
    partially funded by the French ANR projects ANR-16-CE40-0023
    (DESCARTES) and ANR-17-CE40-0015 (DISTANCIA).}}

\author{%
  Marthe Bonamy%
  \thanks{CNRS-LaBRI, University of Bordeaux, France,
    \url{marthe.bonamy@u-bordeaux.fr}.}%
  \and%
  Cyril Gavoille%
  \thanks{LaBRI, University of Bordeaux, France,
    \url{gavoille@labri.fr}.}%
  \and%
  Micha\l{}~Pilipczuk%
  \thanks{Institute of Informatics, University of Warsaw, Poland,
    \url{michal.pilipczuk@mimuw.edu.pl}.}%
}

\begin{titlepage}
\def\thepage{}
\thispagestyle{empty}
\maketitle

\begin{textblock}{20}(0, 12.5)
  \includegraphics[width=40px]{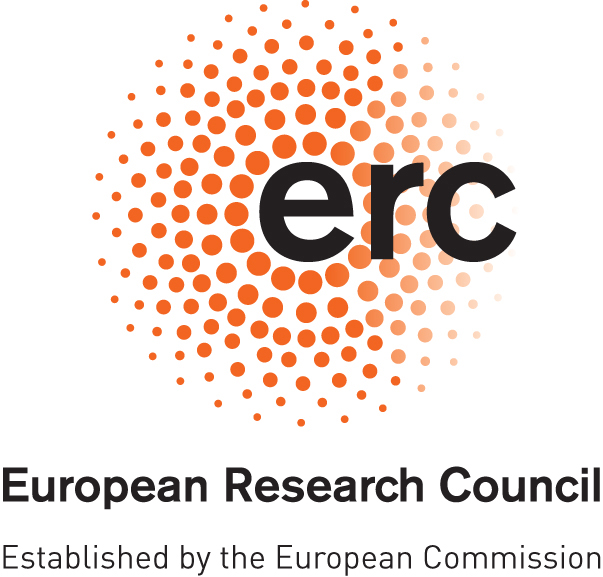}%
\end{textblock}
\begin{textblock}{20}(-0.25, 12.9)
  \includegraphics[width=60px]{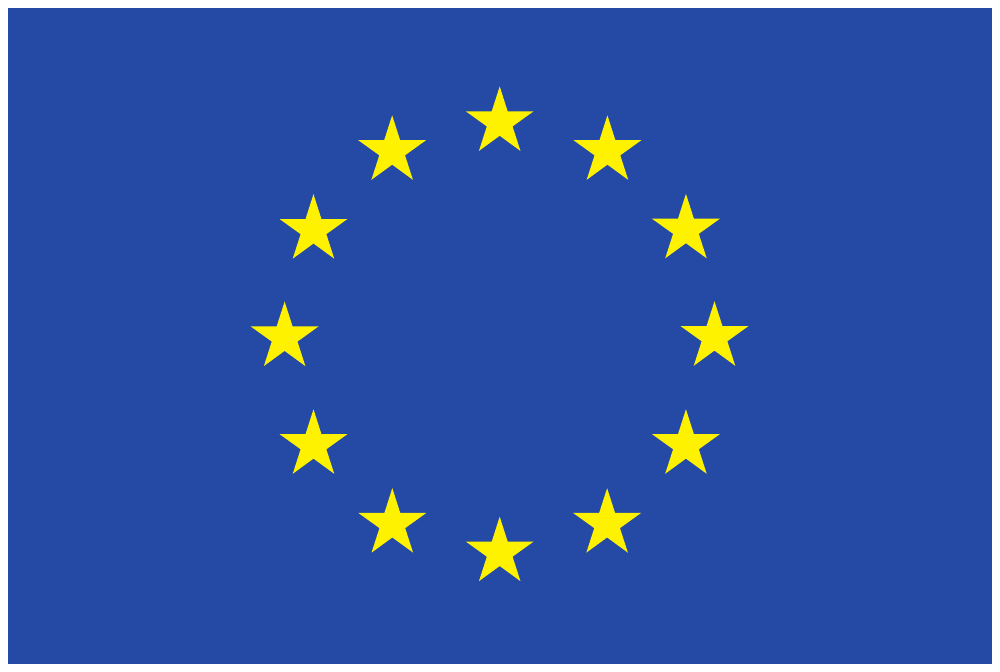}%
\end{textblock}

\begin{abstract}
  An \emph{adjacency labeling scheme} for a given class of graphs is
  an algorithm that for every graph $G$ from the class, assigns bit
  strings (labels) to vertices of $G$ so that for any two vertices
  $u,v$, whether $u$ and $v$ are adjacent can be determined by a fixed
  procedure that examines only their labels. It is known that planar
  graphs with $n$ vertices admit a labeling scheme with labels of bit length
  $(2+o(1))\log{n}$.
  In this work we improve this bound by designing a labeling scheme
  with labels of bit length $(\frac{4}{3}+o(1))\log{n}$. All the labels of the
  input graph can be computed in polynomial time, while adjacency can be
  decided from the labels in constant time.
  
  In graph-theoretical terms, this implies an explicit construction of
  a graph on $n^{4/3+o(1)}$ vertices that contains all planar graphs
  on $n$ vertices as induced subgraphs, improving the previous best
  upper bound of $n^{2+o(1)}$.

  Our labeling scheme can be generalized to larger classes of topologically-constrained graphs, for instance to graphs embeddable in any fixed surface or to $k$-planar graphs for any fixed $k$, at the cost of larger second-order terms.
  
\end{abstract}

\end{titlepage}

\def\path{.}

\section{Introduction}\label{sec:intro}

When representing graphs, say with adjacency lists or
matrices, vertex identifiers usually do not play any particular role with
respect to the structure of the graph: they are essentially just pointers in the
data structure. In contrast, a graph is \emph{implicitly
  represented} when each vertex of the graph is associated to more
information so that adjacency, for instance, can be efficiently
determined from the identifiers without the need of any
global data-structure (cf.~\cite{KNR88,Spinrad03}). 
For example, if
$G$ is an interval graph with $n$ vertices, one can associate with
each vertex $u$ some interval $I(u) \subseteq [1,2n]$ with integer endpoints so
that $u,v$ are adjacent if and only if
$I(u) \cap I(v) \neq \emptyset$. Clearly, no adjacency lists or matrices
are required anymore. Although $G$ may have a quadratic number of
edges, such an implicit representation uses
$2\log{n} + O(1)$ bits per vertex\footnote{Throughout the paper,
  by $\log{n}$ we denote the binary logarithm of $n$.}, regardless of its
degree, which is asymptotically optimal~\cite{GP08}. Compact
representations have several advantages, not only for the memory
storage, but also from algorithmic perspectives. For instance, given a
succinct representation, BFS traversal can be done in $O(n)$
time~\cite{RLDL94,ACJR19}, even if the graph has $\Omega(n^2)$
edges. Speedups due to succinct representations are ubiquitous in
the design of algorithms and data structures.

Formally introduced by Peleg~\cite{Peleg00,Peleg05}, \emph{informative
  labeling schemes} present a way to formalize implicit representations of
graphs. For a given function $\Pi$ defined on pairs of vertices of a
graph from some given class of graphs, an informative labeling scheme
has two components: an {\em{encoding algorithm}} that associates with each vertex a
piece of information (label); and a 
{\em{decoding algorithm}} that computes
$\Pi(u,v,G)$, the value of $\Pi$ applied on vertices $u,v$ of the
graph $G$. The input of the decoding algorithm consists solely of the labels of
$u$ and of $v$, with no other information provided. So, finding an implicit
representation of a graph $G$ can be restated as computing an {\em{adjacency labeling scheme}} for $G$, that is, an informative labeling scheme where $\Pi(u,v,G)$ is \textsc{true} if and only if $u,v$ are
adjacent in $G$.

In this paper we will focus on such adjacency
labeling schemes (referred to as {\em{labeling schemes}} from now on), but many functions
$\Pi$ other than adjacency are of great interest. Among them there are
labeling schemes designed for ancestry~\cite{FK10} and lowest common ancestors
in rooted trees~\cite{AGKR04,AHGL14}, distances~\cite{GU16,GKU16,AGHP16,FGNW17} and
forbidden-set distances~\cite{ACGP16}, compact
routing~\cite{FG01a,TZ01,RT15}, flow problems~\cite{KKKP04}, and many
others. We refer to~\cite{GP03c}, and references therein, for a survey
of informative labeling schemes and their applications in distributed
computing, and also to~\cite{Rotbart16} for a survey on recent
developments in labeling schemes specialized for trees.

\paragraph*{Planar graphs.}
Planar graphs are perhaps one of the most studied class of graphs in this area, due to the wide
variety of their implicit representations. To mention just a few, planar
graphs are contact graphs of circles~\cite{Koebe36}, of 3D
boxes~\cite{Thomassen86}, of triangles~\cite{dFOR94}, and more
recently, of L-shapes~\cite{GIP18}. They also have $1$-string
representations~\cite{CGO10}, and their incidence graphs form posets of dimension
three~\cite{Schnyder89}. Each of these representations leads to a labelling scheme where each vertex can be encoded using a label consisting of $O(\log{n})$ bits, independent of its degree.

The first explicit bound on the label length,
given by Kannan et al.~\cite{KNR88}, was $4\ceil{\log{n}}$ bits. Using the fact that planar graphs
have arboricity at most three together with a labeling scheme for forests with label length $\log n+o(\log n)$, one can achieve also a similar $3\log{n} + o(\log{n})$
upper bound for planar graphs, where the lower-order term $o(\log{n})$ directly depends on the
second-order term of the bound for forests. It was a challenging
question to optimize this second-order for forests. It has been
successively reduced from $O(\log\log{n})$~\cite{Chung90} to
$ O(\log^*{n})$~\cite{AR02b}, and then to a constant only recently by Alstrup
et al.~\cite{ADBTK17}. As explained above, this leads to an upper bound of $3\log{n} + O(1)$ for
planar graphs. By improving the labeling scheme for
bounded treewidth graphs, namely from $O(k\log{n})$~\cite{KNR88} to
$\log{n} + O(k\log\log{n})$, Gavoille and Labourel~\cite{GL07} showed
that partitioning the edges of a planar graph into two bounded
treewidth subgraphs, rather than into three forests, leads to a
shorter representation: with labels consisting of $2\log{n} + O(\log\log{n})$ bits. Until this work, this has been the best known upper bound for planar graphs.

The best known results for several subclasses of planar graphs are reported in
Table~\ref{table:rec}.

\begin{table}[htb!]\label{table:rec}
  \begin{center}
    \renewcommand{\arraystretch}{1.1}
    \def\new{\textbf{[this paper]}}
    \begin{tabular}{ccc}
      \hline\hline
      Graph classes & Upper bound & References\\
      (with $n$ vertices) & (label length in bits) & \\
      \hline\hline\\[-2.5ex]
      maximum degree-$2$ & $\log{n} + O(1)$ & \cite{AAHBx16,Butler09,ELO08}\\
      caterpillars & $\log{n} + O(1)$ & \cite{BGL06}\\
      bounded degree trees & $\log{n} + O(1)$ & \cite{Chung90}\\ 
      bounded depth trees & $\log{n} + O(1)$ & \cite{FK10b}\\ 
      trees & $\log{n} + O(1)$ & \cite{ADBTK17}\\
      bounded degree outerplanar & $\log{n} + O(1)$ & \cite{Chung90,AR14}\\ 
      outerplanar & $\log{n} + O(\log\log{n})$ & ~\cite{GL07}\\
      bounded treewidth planar & $\log{n} + O(\log\log{n})$ & \cite{GL07}\\
      maximum degree-$4$ planar & $\frac{3}{2}\log{n} + O(\log\log{n})$ & \cite{AR14}\\
      bounded degree planar & $2\log{n} + O(1)$ & \cite{Chung90}\\
      planar & $2\log{n} + O(\log\log{n})$ & \cite{GL07}\\[1ex]
      \hline\hline\\[-2.5ex]
      planar & $\frac{4}{3}\log{n} + O(\log\log{n})$ & \new\\[1ex]
      \hline\hline
    \end{tabular}
    \caption{State-of-the-art for adjacency labeling schemes on planar
      graphs and some subclasses. The bounds from
      references~\cite{Chung90,ELO08,Butler09} come from
      induced-universal graphs, whereas all the others come from
      labeling schemes. The only known lower bound for planar graphs
      is $\log{n} + \Omega(1)$.}
  \end{center}
\end{table}

\paragraph*{Our contribution.}
In this work we present a new labeling scheme for planar graphs that uses labels of length bounded\footnote{For brevity, in this informal exposition we ignore  additive terms of lower order $o(\log n)$.} by $\frac{4}{3}\log n$. Note that this not only improves the previously best known bound of $2\log n$ for general planar graphs~\cite{GL07}, but even the refined bound of $\frac{3}{2}\log n$ for the case of planar graphs of maximum degree $4$~\cite{AR14}.

The main ingredient of our result is the recent product structure theorem of Dujmovi\'{c} et al.~\cite{DJMPx19},
which says the following: 
Every planar graph $G$ is a subgraph of a graph of the form $H\stimes P$, where $H$ is a graph of treewidth at most $8$, $P$ is a path, and $\stimes$ denotes the {\em{strong graph product}} (see Section~\ref{sec:prelims} for a definition). Moreover, $H$, $P$, and a subgraph embedding witnessing this can be found in polynomial time.

The first step in our proof is the design of an auxiliary labeling scheme with labels of length $\log{n} + \log{d}$, assuming that the graph $G$ in question is given together with an embedding into $H\stimes P$, where $H$ has bounded treewidth and $P$ is a path of length $d$. This parameterized bound is never worse
than the currently best known bound of $2\log{n}$, because we may always assume $d<n$, but later in the general case we use it for $d=O(n^{1/3})$.
We remark that the proof of the product structure theorem of Dujmovi\'{c} et al.~\cite{DJMPx19} in fact yields a subgraph embedding into $H\stimes P$ where $P$ has length bounded by the diameter of the considered graph $G$, so as a side result we obtain a labeling scheme for planar graphs with diameter $d$ that uses labels of  length bounded by $\log{n}+\log{d}$.

The second step --- the main case --- relies on the layering technique applied on the structure provided by the product structure theorem.
Precisely, for a given planar graph $G$ we compute a subgraph embedding $\phi$ of $G$ into $H\stimes P$, where $H$ is a graph of bounded treewidth and $P$ is a path (with no nontrivial bound on its length).
We choose a parameter $d\geq 3$ (which will be set later) and divide $H\stimes P$ into blocks of width $d$; that is, each block is of the form $H\stimes Q$ where $Q$ is a subpath of $P$ consisting of $d$ consecutive vertices.
By mapping the blocks through $\varphi^{-1}$ back to $G$, we thus divide $G$ into {\em{strips}}, where each strip can be embedded into $H\stimes Q$ where $Q$ has length $d-1$.
These strips are separated by {\em{borders}} whose union is a graph on $O(n/d)=O(n^{2/3})$ vertices and of constant treewidth.
Using the known bounds for graphs of bounded treewidth~\cite{GL07}, for this border graph we can compute a labeling $\lambda_1$ with labels of length $\log{(n/d)}$. 
On the other hand, to the union of strips we can apply the auxiliary labeling scheme explained in the previous paragraph, and thus obtain a labeling $\lambda_2$ for the strips with labels of length $\log{n}+\log d$.

At this point,
superposing the two schemes $\lambda_1$ and $\lambda_2$ would give a labeling scheme of length $2\log n$. This is because vertices appearing at the borders of strips have to inherit labels from both labelings: $\log{(n/d)}$ from $\lambda_1$ and $\log{n}+\log{d}$ from $\lambda_2$, which sums up to $2\log{n}$. So far, this yields no improvement over the previous results.
However, by revisiting the scheme for graphs of bounded
treewidth we are able to show that for vertices at the borders --- whose number is $O(n/d)$ --- the labeling $\lambda_2$ can use much shorter labels: only of length
$\log{(n/d)}$ instead of $\log{n} + \log{d}$. Hence, the combined labels of border vertices are of length at most $2\log{(n/d)}$, implying that every vertex receives a label of length bounded by
\[
  \max\set{ ~\log{n} + \log{d}~ , ~2\log{(n/d)}}.
\]
This expression is minimized for $d = n^{1/3}$ and then evaluates to
$\frac{4}{3}\log{n}$, the desired bound.

Finally, we observe that the only property implied by planarity that we used in our labeling scheme is the product structure given by the theorem of Dujmovi\'{c} et al.~\cite{DJMPx19}.
Precisely, if we assume that we work with a class of graph $\Cc$ such that every graph $G\in \Cc$ admits a polynomial-time computable subgraph embedding into a graph $H\stimes P$, where $H$ has constant treewidth
and $P$ is a path, then the whole reasoning goes through. We call such graph classes {\em{efficiently flat}} and choose to work throughout the paper with this abstract property alone, instead of the concrete case of planar graphs. The reason for this is that following the result of Dujmovi\'c{} et al.~\cite{DJMPx19} for planar graphs, many more general classes of graphs have been rendered efficiently flat, for instance graphs embeddable into any fixed surface~\cite{DJMPx19}, or $k$-planar graphs for any fixed $k$~\cite{DMW19} (see Section~\ref{sec:prelims} for more examples). Consequently, our result gives a labeling scheme of length $\frac{4}{3}\log n$ for all these classes.

In all our labeling schemes, given the input graph we can compute the labeling of its vertices in polynomial time, while the adjacency can be determined from the labels in constant time.

\paragraph*{Connections with universal graphs.}
It has been observed in~\cite{KNR88} that the design of labeling schemes
with short labels is tightly connected with the construction of small
induced-universal graphs. Recall that a graph $\sU$ is induced-universal for a
given set of graphs $\sS$ if every graph $G\in \sS$ is isomorphic
to some induced subgraph of $\sU$. Then graphs from $\sS$ admit a labeling scheme
with $k$-bit labels if and only if $\sS$ has an induced-universal
graph $\sU$ with at most $2^k$ vertices, see~\cite{KNR88}.
Thus, our labeling scheme provides an explicit construction of an induced-universal graph for $n$-vertex planar graphs that has $n^{4/3+o(1)}$ vertices, improving upon the previously best known bound of $n^{2+o(1)}$, derived from~\cite{GL07}.

The search for optimum bounds on the sizes of induced-universal graphs is a well-studied topic, see for example the recent developments for general $n$-vertex graphs~\cite{Alon17,AKTZ15} and for $n$-vertex trees~\cite{ADBTK17}. We refer readers interested in this topic to the recent survey of Alstrup et al.~\cite{AKTZ19}.

Apart from induced-universal graphs, there is also an alternative definition: {\em{edge-universal graphs}}. Here, we say that $\sU$ is {\em{edge-universal}} for a set of graph $\sS$ if every graph from $\sS$ is a subgraph of $\sU$ (not necessarily induced). As far as edge-universal graphs for $n$-vertex planar graphs are concerned, there are much more concise constructions than in the induced setting. Babai et al.~\cite{BCEGS82} gave a construction with $O(n^{3/2})$ edges, while if one restricts the question to $n$-vertex planar graphs with constant maximum degree, then the number of edges can be reduced even to $O(n)$~\cite{Capalbo02}. However, in general it is unclear how edge-universal graphs can be turned into induced-universal graphs without a significant explosion in the size, see e.g. the discussion in~\cite{Chung90}.

\paragraph*{Subsequent work.} After the publication of the conference version of this work~\cite{BGP20}, Dujmovi\'c{} et al.~\cite{DEJGMM20} announced a construction of a labeling scheme of optimum length $(1+o(1))\log n$ for any efficiently flat class of graphs. Their proof is also based on the product structure, but is much more involved. Moreover, in the scheme of~\cite{DEJGMM20}, the adjacency tests take time $O(\sqrt{\log n\log \log n})$, as compared to constant time for our scheme.

In the conference version of this work~\cite{BGP20} we concentrated on the case of planar graphs and, more generally, graphs embeddable into any fixed surface. Instead of using the product structure abstractly, we relied on a more hands-on combinatorial understanding via BFS layerings and partitions with bounded-treewidth quotient graphs. In particular, in several places we relied on auxiliary properties of the considered classes, like being minor-closed. The version presented here relies on the product structure alone and thus is more general.

\paragraph*{Organization.} In Section~\ref{sec:prelims} we recall the main definitions and results regarding labeling schemes, tree decompositions, and the product structure theorem of Dujmovi\'c{} et al.~\cite{DJMPx19}. Then in Section~\ref{sec:tw} we revisit and strengthen the labeling scheme for graphs of bounded treewidth of Gavoille and Labourel~\cite{GL07}. In Section~\ref{sec:small-diam} we give an auxiliary scheme for graphs for which the product structure theorem yields an embedding into the strong product of a bounded treewidth graph and a {\em{short}} path. This result is then used in Section~\ref{sec:planar} to treat the general case of graphs from an efficiently flat class. We conclude in Section~\ref{sec:conc} by discussing some further research directions.

\section{Preliminaries}\label{sec:prelims}

We use standard graph notation. For a graph $G$, the vertex and edge sets of $G$ are denoted by $V(G)$ and $E(G)$, respectively. For $A\subseteq V(G)$, we write $G[A]$ for the subgraph of $G$ induced by $A$ and $G-A$ for the subgraph of $G$ induced by $V(G)\setminus A$. A {\em{subgraph embedding}} of a graph $H$ into a graph $G$ is an injective function $\phi\colon V(H)\to V(G)$ such that $uv\in E(H)$ entails $\phi(u)\phi(v)\in E(G)$.


\paragraph*{Labeling schemes.} The following definition formalizes the concept of labeling schemes.

\begin{definition}
  Let $\sC$ be a class of graphs. An \emph{adjacency labeling scheme}
  for $\sC$ is a pair $\langle\lambda,\xi\rangle$ of functions such
  that, for every graph $G \in \sC$, it holds:

  \begin{itemize}[noitemsep,topsep=0ex]

  \item $\lambda$ is the \emph{Encoder} that assigns to every vertex
    $u$ of $G$ a different binary string $\lambda(u,G)$; and

  \item $\xi$ is the \emph{Decoder} that decides adjacency from the
    labels taken from $G$. More precisely, for every pair $u,v$ of
    vertices of $G$, $\xi(\lambda(u,G), \lambda(v,G))$ is
    \textsc{true} if and only if $u,v$ are adjacent in $G$.

  \end{itemize}

  The \emph{length} of the labeling scheme $\langle \lambda,\xi\rangle$ is the function $\ell\colon \N\to \N$ that maps every $n\in \N$ to the maximum length, expressed in the number of bits, of  labels assigned by the Encoder in $n$-vertex graphs from $\sC$.
\end{definition}

In the above definition we measure the length only in terms of the vertex count $n$, but we can extend the definition to incorporate auxiliary graph parameters, like diameter or treewidth, in a natural way.
Whenever $G$ is clear from the context, we write $\lambda(u)$ as a shorthand for
shorthand for $\lambda(u,G)$.

When speaking about the complexity of Encoder and Decoder, we assume
RAM model with machine words of bit length $O(\log{n})$ and unit cost
arithmetic operations.

\paragraph*{Tree decompositions.}
A \emph{tree decomposition} of a graph $G$ is a pair $(T,\beta)$, where $T$ is a tree and $\beta$ maps every node $x$ of $T$ to its \emph{bag} $\beta(x)\subseteq V(G)$ so that: for every edge $uv$ of $G$ there exists a node $x$ satisfying $\set{u,v}\subseteq \beta(x)$, and for every vertex $u$ of $G$, the set $\set{x\in V(T)\colon u\in \beta(x)}$ induces a non-empty, connected subtree of $T$. The \emph{width} of $(T,\beta)$ is
$\max_{x\in V(T)} {|\beta(x)|-1}$, while the \emph{treewidth} of $G$
is the minimum possible width of a tree decomposition of~$G$.

\paragraph*{Flatness.}
For two graphs $G$ and $H$, the {\em{strong product}} of $G$ and $H$, denoted $G\stimes H$,
is the graph on vertex set $V(G)\times V(H)$ where two different vertices $(u,v)$ and $(u',v')$ are adjacent if and only if vertices $u$ and $u'$ are equal or adjacent in $G$, and vertices $v$ and $v'$ are equal or adjacent in $H$. The following definition describes the key structural property discovered by Dujmovi\'c{} et al.~\cite{DJMPx19}.

\begin{definition}
A class of graph $\Cc$ is {\em{flat}} if there exists $w\in \N$ such that every graph $G\in \Cc$ is a subgraph of some graph of the form $H\stimes P$, where $H$ has treewidth at most $w$ and $P$ is a path. 
\end{definition}

Note that in the above definition one may assume that $|V(H)|\leq |V(G)|$, as one can remove every vertex $v$ of $H$ such that no element of the fiber $\{(v,i)\colon i\in V(P)\}$ participates in the subgraph embedding of $G$ into $H\stimes P$. Similarly, we may assume that $|V(P)|\leq |V(G)|$.

As proved by Dujmovi\'c{} et al.~\cite{DJMPx19}, planar graphs are flat. However, this property carries over to more general classes of topologically-constrained graphs, as the following classes are flat as well:
\begin{itemize}[nosep]
    \item graphs of Euler genus $g$, for every fixed $g\in \N$~\cite{DJMPx19};
    \item every apex-minor-free class~\cite{DJMPx19};
    \item every proper minor-closed class with bounded maximum degree~\cite{DJMPx19};
    \item $(0,g,k,p)$-nearly embeddable graphs, for all fixed $g,k,p\in \N$~\cite{DJMPx19};
    \item $k$-planar graphs, for every fixed $k$~\cite{DMW19}.
\end{itemize}
See also~\cite{DMW19} for several other examples of flat classes, and~\cite{DHJLW20} for a survey of the area.

In our proofs we will need to assume algorithmic aspects of flatness. Precisely, we shall say that a flat class of graph $\Cc$ is {\em{efficiently flat}} if given $G\in \Cc$, one can in polynomial time compute a graph $H$ of treewidth at most $w$, for a constant $w\in \N$, a path $P$, and a subgraph embedding of $G$ into $H\stimes P$. Fortunately, a close inspection of the proofs in~\cite{DJMPx19} shows that all the abovementioned flat classes are actually efficiently flat, so our results apply to all of them.

Throughout the paper we will focus on proving the following result, from which all the corollaries discussed in Section~\ref{sec:intro} follow.

\begin{theorem}\label{thm:main}
  Every efficiently flat class of graphs admits a labeling scheme of length $\frac{4}{3}\log n+O(\log \log n)$. The Encoder runs in polynomial time and the Decoder in constant time.
\end{theorem}

\section{Bounded Treewidth Graphs}\label{sec:tw}

Like the construction of~\cite{GL07} for planar graphs, our result
relies on the labeling scheme developed for bounded treewidth graphs.

\begin{theorem}[\cite{GL07}]\label{thm:bnd-tw}
For any fixed $k\in \N$, graphs of treewidth at most $k$ admit a labeling scheme of length $\log{n} + O(k\log\log{n})$. The Encoder runs in $O(n\log{n})$ time and the Decoder runs in constant time.
\end{theorem}

In later sections we significantly rely on the combinatorics behind
the proof of Theorem~\ref{thm:bnd-tw}. We will need two ingredients:
\begin{enumerate}[label=(\arabic*),ref=(\arabic*),noitemsep,topsep=0pt]
\item an understanding of how encoding and decoding works in the
  labeling scheme; and
\item a strengthening of the result, where we can assume that a
  prescribed set of at most $q$ vertices receives shorter labels,
  namely of length $\log{q} + O(k\log\log{n})$.
\end{enumerate} 
These two properties are formally stated as follows.

\begin{theorem}\label{thm:stronger-tw}
  For any fixed $k\in \N$, the class of graphs of treewidth at most $k$ admits a labeling scheme $\langle \lambda,\phi\rangle$ of length $\log n+O(k\log \log n)$ with the following properties:
  \begin{enumerate}[label=(P\arabic*),ref=(P\arabic*),noitemsep,topsep=0pt]
      \item\label{p:ids} From any label $a$ one can extract in time $O(1)$ an identifier $\iota(a)$, so that the Decoder may be implemented as follows: given a label $a$, one may compute in
  time $O(k)$ a set $\Gamma(a)$ consisting of at most $k$ identifiers so
  that $\phi(a,b)$ is \textsc{true} if and only if 
  $\iota(a)\in \Gamma(b)$ or
  $\iota(b)\in \Gamma(a)$.
      \item\label{p:saving} If the input graph $G$ is given together with a vertex subset $Q$,
  then the scheme can assign to the vertices of $Q$ labels of length
  $\log{|Q|} + O(k\log\log{n})$.
  \end{enumerate}
  The Encoder works in time $O(n \log n)$ while the Decoder works in constant time.
\end{theorem}


The proof of Theorem~\ref{thm:stronger-tw} largely follows the approach of Gavoille and Labourel~\cite{GL07}. In particular, their scheme achieves property~\ref{p:ids} without any modifications. However, to achieve property~\ref{p:saving} we need to replace a crucial combinatorial element of the proof with a new argument.

The remainder of this section is devoted to the presentation of the proof of  Theorem~\ref{thm:stronger-tw}, which largely follows the approach of Gavoille and Labourel~\cite{GL07}. In Section~\ref{sec:encoding} we recall this approach and explain that property~\ref{p:ids} follows from it without any modifications. In Section~\ref{sec:saving} we replace a crucial ingredient of~\cite{GL07} with a new argument in order to achieve property~\ref{p:saving} as well.

\subsection{Encoding and decoding}\label{sec:encoding}

We start with a brief presentation of the approach of Gavoille and Labourel~\cite{GL07}.
Our presentation is a bit
simplified compared to that of~\cite{GL07}, because we choose not to
optimize the label length as much as there (e.g. Gavoille and Labourel
actually provide an upper bound of $\log{n} + O(k\log\log{(n/k)})$
instead of $\log{n} + O(k\log\log{n})$ by a more precise analysis).

First, since the input graph $G$ has treewidth at most~$k$, one can
obtain a chordal supergraph $G^+$ of $G$ on the same vertex set such
that $G^+$ also has treewidth at most $k$. This can be done as
follows: take a tree decomposition of $G$ of width at most $k$ and
turn every bag into a clique. Since for fixed $k$ such a tree decomposition can be computed in linear time~\cite{Bod96a}, $G^+$ can
be computed in linear time.

Next, it is well-known that since $G^+$ is chordal and of treewidth at
most $k$, in linear time we can compute an orientation $\vec{G}$ of
$G^+$ such that every vertex $u$ has at most $k$ out-neighbors in
$\vec{G}$, and moreover $u$ together with those out-neighbors form a
clique in $G^+$. For every $u\in V(G)$, let $K_u$ be the set
consisting of $u$ and its out-neighbors in $\vec{G}$.

The key idea of the approach of Gavoille and Labourel is to compute a
\emph{bidecomposition} of the graph $G^+$, which is a notion roughly
resembling tree decompositions, but actually quite different.

%

\begin{definition}
  A \emph{bidecomposition} of a graph $H$ is a pair $(T,\alpha)$,
  where $T$ is a binary rooted tree and $\alpha$ maps vertices $H$ to
  nodes of $T$, so that for every edge $uv$ of $H$, $\alpha(u)$ and
  $\alpha(v)$ are related.
\end{definition}

As proved in~\cite{GL07}, graphs of bounded treewidth admit
bidecompositions with small parts. This is the key combinatorial
ingredient of the proof.

\begin{lemma}[cf. Lemma~1 in~\cite{GL07}]\label{lem:bidecomposition}
  Let $G$ be an $n$-vertex graph of treewidth at most $k$. Then there
  exists a bidecomposition $(T,\alpha)$ of $G$ satisfying the
  following:
\begin{enumerate}[label=(A\arabic*),ref=(A\arabic*),noitemsep,topsep=0pt]
\item\label{pa:parts} $|\alpha^{-1}(x)| = O(k\log{n})$ for every
  node $x$ of $T$; and
\item\label{pa:depth} $T$ has depth at most $\log{n}$.
\end{enumerate}
Moreover, for every fixed $k$, given $G$ such a bidecomposition can be
constructed in time $O(n\log{n})$.
\end{lemma}

We apply Lemma~\ref{lem:bidecomposition} to the graph $G^+$, thus
getting a suitable bidecomposition $(T,\alpha)$. Based on this, a
labeling is constructed as follows.

Consider any $u\in V(G)$. Since $K_u$ is a clique in $G^+$, it follows
that nodes $\set{\alpha(v)}_{v\in K_u}$ are pairwise related. Hence,
there exists a path $P_u$ in $T$ starting at the root that contains
all nodes $\alpha(v)$ for $v\in K_u$. The second endpoint of $P_u$ is
the deepest among nodes $\set{\alpha(v)}_{v\in K_u}$. Let $P'_u$ be
the prefix of $P_u$ from the root of $T$ to $\alpha(u)$.

For each node $x$ of $T$ fix an arbitrary enumeration of
$\alpha^{-1}(x)$ using index taken from $[0,|\alpha^{-1}(x)|)$.  Now,
the identifier of vertex $u$ consists of the following pieces of
information:
\begin{enumerate}[noitemsep,topsep=0pt]
\item The encoding of the path $P'_u$ as a bit string of length
  $|V(P'_u)|-1$ that encodes, for consecutive non-root vertices of
  $P'_u$, whether they are left or right children.
\item The index of $u$ within $\alpha^{-1}(\alpha(u))$.
\item The depth of $\alpha(u)$ in $T$.
\end{enumerate}
Since $T$ has depth at most $\log{n}$ and
$|\alpha^{-1}(x)| = O(k\log{n})$ for every node $x$ of $T$, we
conclude that the identifier has total length
$\log{n} + \log{k} + O(\log\log{n})$.  In addition to the identifier,
the label of $u$ contains the following pieces of information:
\begin{enumerate}[noitemsep,topsep=0pt]
\item Encoding of the suffix of $P_u$ that is not contained in $P'_u$;
  this, together with the information from the identifier, adds up to
  the encoding of $P_u$.
\item For every $v\in K_u\setminus \set{u}$, the depth of $\alpha(v)$ in
  $T$, the index of $v$ within $\alpha^{-1}(\alpha(v))$, and whether
  the edge $uv$ belongs to $E(G)$ (it may belong to
  $E(G^+)\setminus E(G)$).
\end{enumerate}
As shown in~\cite{GL07}, the above information, together with the
identifier, can be encoded in $\log{n}+O(k\log\log{n})$ bits,
resulting in the promised upper bound on the label length.
Moreover, given the bidecomposition $(T,\alpha)$ the labeling can be computed in linear time, assuming $k$ is fixed.

It is now straightforward to see that from the label of $u$ one can
derive the identifiers of the out-neighbors of $u$ in $G^+$. Indeed,
for every $v\in K_u\setminus\set{u}$ the depth of $\alpha(v)$ and the
index of $v$ in $\alpha^{-1}(\alpha(v))$ are directly stored in the
label of $u$, while the encoding of the path $P'_v$ can be obtained by
taking the encoding of $P_u$ and trimming it to the prefix of length
equal to the depth of $\alpha(v)$. With every such out-neighbor $v$ we
have also stored the information whether the edge $uv$ is contained in
$G$, or it was added when modifying $G$ to $G^+$. Hence, given the
label $\lambda(u)$ we can compute a set of at most $k$ identifiers of
neighbors of $u$, which is a suitable set $\Gamma(\lambda(u))$. This
proves property~\ref{p:ids}.

\subsection{Saving on labels of a small set of vertices}
\label{sec:saving}

We now explain how the general approach of Gavoille and Labourel~\cite{GL07}, presented in the previous section, can be amended to achieve property~\ref{p:saving} as well.
The difference is that we
replace the usage of Lemma~\ref{lem:bidecomposition} with the
following Lemma~\ref{lem:bidecomposition-stronger}.

\begin{lemma}\label{lem:bidecomposition-stronger}
  Let $G$ be an $n$-vertex graph of treewidth at most $k$ and
  $S\subseteq V(G)$. Then there exists a bidecomposition $(T,\alpha)$
  of $G$ satisfying the following:
\begin{enumerate}[label=(B\arabic*),ref=(B\arabic*),noitemsep,topsep=0pt]
\item\label{p:parts} $|\alpha^{-1}(x)| = O(k\log{n})$ for every node
  $x$ of $T$;
\item\label{p:depth} $T$ has depth at most $\log{n} + O(1)$; and
\item\label{p:Sdepth} for every $u\in S$, $\alpha(u)$ is at depth at
  most $\log{|S|} + O(1)$ in $T$.
\end{enumerate}
Moreover, for every fixed $k$, given $G$ and $S$ such a decomposition
can be constructed in time $O(n \log n)$.
\end{lemma}

Consider the set $Q$ of prescribed vertices as in
property~\ref{p:saving}, and apply
Lemma~\ref{lem:bidecomposition-stronger} to $G^+$ with
\[
  S =\bigcup_{u\in Q} K_u.
\]
We have $|S| \le (k+1)\cdot |Q|$.  Hence, in the notation of the
previous section, for every $u\in Q$ we have that $P_u$ has at most
$\log{|S|} + O(1) = \log{|Q|} + O(\log{k})$ nodes, while for every
other vertex $u$ we have that $P_u$ has at most $\log{n} + O(1)$
nodes. Plugging this into the analysis of the previous section gives
the desired bounds on the lengths of labels in the constructed
labeling. Note that thus, property~\ref{p:ids} still holds, while property~\ref{p:saving} is achieved.

We are left with proving Lemma~\ref{lem:bidecomposition-stronger}. We
would like to stress that this is \emph{not} a simple modification of
the proof of Lemma~\ref{lem:bidecomposition} presented
in~\cite{GL07}. The general idea is to recursively decompose the
graph, where at each step we use a separator of size $O(k\log{n})$ to
split the graph into two parts, each containing (roughly) at most half
of the remaining vertices and at most half of the remaining vertices
of $S$. In~\cite{GL07} only the first objective --- halving the total
number of vertices --- was necessary, and this was relatively easy to
achieve using a separator of size $O(k\log{n})$. However, the strategy
used in~\cite{GL07} does not generalize to achieving both objectives
at the same time. Hence, our splitting step is based on a
different argument.

\begin{proof}[of Lemma~\ref{lem:bidecomposition-stronger}]
  We first focus on proving the existential statement. Then, at the end, we discuss the
  algorithmic aspects of the proof.

  For a graph $H$ and a nonnegative weight function $\wei\colon V(H)\to \R_{\geq 0}$,
  we write $\wei(H) = \sum_{u\in V(H)} \wei(u)$. 
  We first need a robust understanding of balanced separators in graphs of bounded treewidth, which boils down to a understanding balanced separators in trees. We will use the following well-known claim.
  
  \begin{claim}\label{cl:balanced-trees}
    Let $\eps>0$, let $T$ be a tree, and let $\wei\colon V(T)\to\ \R_{\geq 0}$ be a nonnegative weight function on the nodes of $T$. Then there exists a set of nodes $S\subseteq V(T)$ of size at most $\eps^{-1}$ such that for every connected component~$C$ of $T-S$, we have $\wei(C)\leq \eps\cdot \wei(T)$. 
  \end{claim}
  \begin{clproof}
    The claim is trivial if $\wei(T)=0$, so assume that $\wei(T)>0$. Root $T$ in any node $r$; this induces a natural ancestor/descendant relation. Starting with $S=\emptyset$ and all vertices of $S$ unmarked, we apply the following iterative procedure. As long as the total weight of unmarked vertices is at least $\eps\cdot \wei(T)$, find the deepest node $x$ such that the total weight of unmarked descendants of $x$ (including $x$ itself) is at least $\eps\cdot \wei(T)$, add $x$ to $S$, and mark $x$ together with all its descendants.
    
    As with every node added to $S$ we mark nodes of total weight at least $\eps\cdot \wei(T)$, the final set $S$ has size at most $\eps^{-1}$. The fact that every connected component of $T-S$ has weight at most $\eps\cdot \wei(T)$ follows easily from always choosing $x$ as the deepest node with the considered property.
  \end{clproof}
  
  Claim~\ref{cl:balanced-trees} can be generalized to graphs of bounded treewidth as follows.
  
  \newcommand{\bwei}{\wei'}
  
  \begin{claim}\label{cl:epsseps}
  Let $\eps>0$, let $H$ be a graph of treewidth at most $k$, and let
  $\wei\colon V(H)\to \R_{\geq 0}$ be a nonnegative weight function on the vertices of
  $H$.  Then there exists a set $Z\subseteq V(H)$ of size
  at most $\eps^{-1}\cdot (k+1)$ such that for every connected component $C$ of $H-Z$
  we have $\wei(C)\leq \eps\cdot\wei(H)$.
  \end{claim}
  \begin{clproof}
    Let $(T,\beta)$ be a tree decomposition of $H$ of width at most $k$.
    Let us root $T$ in any node $r$; this induces a natural parent/child relation in $T$.
    For a node $x$ of $T$, we define the {\em{margin}} of $x$ as
    $$\mu(x)=\beta(x)\setminus \beta(y),$$
    where $y$ is the parent of $x$ in $T$. For the root $r$ we set $\mu(r)=\beta(r)$.
    For a node $x\in T$, we define
    $$\bwei(x) = \wei(\mu(x)).$$
    Thus, $\bwei\colon V(T)\to \R_{\geq 0}$ is a nonnegative weight function defined on the nodes of $T$.
    By observing that $\{\mu(x)\colon x\in V(T)\}$ is a partition of $V(H)$, we see that $\bwei(T)=\wei(H)$.
    
    By applying Claim~\ref{cl:balanced-trees} to tree $T$ with weight function $\bwei$, we obtain a suitable node subset $S\subseteq V(T)$ of size at most $\eps^{-1}$.
    Let $Z=\bigcup_{x\in S} \beta(x)$; then $|Z|\leq \eps^{-1}\cdot (k+1)$.
    To see that each connected component~$C$ of $H-Z$ satisfies $\wei(C)\leq \eps\cdot \wei(H)$, 
    note that such a component must be entirely contained in $\bigcup_{x\in V(D)} \mu(x)$ for some connected component $D$ of $T-S$, 
    while $\bwei(D)\leq \eps\cdot \bwei(T)=\eps\cdot \wei(H)$ by Claim~\ref{cl:balanced-trees}.
  \end{clproof}
  
We now use Claim~\ref{cl:epsseps} to find separators that are suited
for constructing a bidecomposition. The idea is that after removing
the separator, we need to be able to group the remaining components
into two parts that are roughly balanced: both in terms of the number
of vertices and in terms of the number of vertices of $Q$. We prefer
to put this condition in an abstract way, using two weight functions.
Precisely, we will use the following auxiliary claim.

\begin{claim}\label{cl:balsep-abstract}
  Let $\Omega$ be a finite set and $\wei_1,\wei_2\colon \Omega\to \R_{\geq 0}$ be two nonnegative weight functions on $\Omega$ such that for every element $x\in \Omega$, we have $\wei_1(x)\leq \eps\cdot \wei_1(\Omega)$ and $\wei_2(x)\leq \eps\cdot \wei_2(\Omega)$, for some $\eps>0$. Then there exists a partition of $\Omega$ into $Y$ and $Z$ such that $\wei_t(W)\leq (1/2+3\eps)\cdot \wei_t(\Omega)$, for all $W\in \{Y,Z\}$ and $t \in\set{1,2}$.
\end{claim}
\begin{clproof}
  If $\wei_1(\Omega)=0$ then the claim follows easily by packing elements into $Y$ greedily until $\wei_2(Y)\geq \wei_2(\Omega)/2$, and defining $Z=\Omega\setminus Y$. A symmetric reasoning can be applied when $\wei_2(\Omega)=0$. Therefore, we are left with the case when $\wei_1(\Omega)>0$ and $\wei_2(\Omega)>0$, so by rescaling the weights we may assume that $\wei_1(\Omega)=\wei_2(\Omega)=1$.
  
  For $x\in \Omega$, let
  $$\xi(x)= \wei_1(x)-\wei_2(x).$$
  Note that
  $$\sum_{x\in \Omega} \xi(x)=0\qquad\textrm{ and }\qquad |\xi(x)|\leq 2\eps \quad\textrm{for every }x\in \Omega.$$
  We now inductively define an ordering $x_1,\ldots,x_n$ of the elements of $\Omega$ as follows. 
  Assuming $x_1,\ldots,x_{i-1}$ has already been defined, we select:
  \begin{itemize}
      \item $x_i$ to be an arbitrary non-positive element of $\Omega\setminus \{x_1,\ldots,x_{i-1}\}$, in case $\sum_{j=0}^{i-1} \xi(x_j)\geq 0$; and
      \item $x_i$ to be an arbitrary positive element of $\Omega\setminus \{x_1,\ldots,x_{i-1}\}$, in case $\sum_{j=0}^{i-1} \xi(x_j)<0$.
  \end{itemize}
  The existence of an element $x_i$ as above is implied by the fact that $\sum_{x\in \Omega} \xi(x)=0$.
  As $|\xi(x)|\leq 2\eps$ for every $x\in \Omega$, the following assertion can be proved by a straightforward induction:
  \begin{equation}\label{eq:beaver}
  \left|\sum_{j=0}^i \xi(x_j)\right|\leq 2\eps\quad\textrm{for every }i\in \{0,1,\ldots,n\}.
  \end{equation}
  
  Let $p\in \{1,\ldots,n\}$ be the smallest index such that $\sum_{j=0}^p \wei_1(x_i)\geq \frac{1}{2}$. As $\wei_1(x_p)\leq \eps$, we also have $\sum_{j=0}^p \wei_1(x_i)\leq \frac{1}{2}+\eps$. Thus, if we define
  $$Y\coloneqq \{x_1,\ldots,x_p\}\qquad\textrm{and}\qquad Z\coloneqq \{x_{p+1},\ldots,x_{n}\},$$
  then we have $1/2\leq \wei_1(Y)\leq 1/2+\eps$, implying $\wei_1(Z)\leq 1/2$. Further, from~\eqref{eq:beaver} we conclude that
  $$2\eps\geq \left|\sum_{j=0}^p \xi(x_j)\right|=|\wei_1(Y)-\wei_2(Y)|,$$
  so $1/2-2\eps\leq \wei_2(Y)\leq 1/2+3\eps$. As $Z=\Omega\setminus Y$, the left inequality implies that $\wei_2(Z)\leq 1/2+2\eps$, so all the required properties are fulfilled.
\end{clproof}

The next statement applies Claim~\ref{cl:balsep-abstract} in the context of a single step of constructing a bidecomposition.

\begin{claim}\label{cl:balsep}
  Let $\eps>0$, let $H$ be a graph of treewidth at most $k$, and let
  $\wei_1,\wei_2\colon V(H)\to \R_{\geq 0}$ be two weight functions on the
  vertices of $H$.  Then there exists a partition $(A,X,B)$ of $V(G)$
  with the following properties:
  \begin{itemize}[noitemsep,topsep=0pt]
  \item there is no edge with one endpoint in $A$ and second in $B$;
  \item $|X| \leq  \eps^{-1}\cdot (k+1)$; and
  \item $\wei_t(A)\leq (1/2+\eps)\cdot \wei_t(H)$ and
    $\wei_t(B)\leq (1/2+\eps)\cdot \wei_t(H)$, for $t \in\set{1,2}$.
  \end{itemize}
\end{claim}
\begin{clproof}
  Apply Claim~\ref{cl:epsseps} to $H$ with weight functions
  $\wei_1(\cdot)$ and $\wei_2(\cdot)$, yielding suitable separators
  $Z_1,Z_2\subseteq V(H)$. Let $X=Z_1\cup Z_2$. Then
  $|X|\leq  \eps^{-1}\cdot (k+1)$ and
  \[
    \wei_t(C)\leq \eps\cdot \wei_t(H)\quad \textrm{for each }t\in
    \set{1,2}\textrm{ and every connected component }C\textrm{ of }H-X.
  \]
  It therefore suffices to prove that the connected components of
  $H-X$ can be partitioned into two groups $A$ and $B$ so that the
  third condition is satisfied.
  
  Let $\Omega$ be the set of connected components of $H-X$. By applying Claim~\ref{cl:balsep-abstract} to $\Omega$ with weight functions $\wei_1,\wei_2$ (treated as weight functions on $\Omega$), we find a partition of $\Omega$ into $Y$ and $Z$ such that $$\wei_t(W)\leq (1/2+3\eps)\cdot \wei_t(\Omega)\leq (1/2+3\eps)\cdot \wei_t(H),$$
  for all $W\in \{Y,Z\}$ and $t\in \{1,2\}$.
  It now remains to define $A=\bigcup_{C\in Y} V(C)$ and $B=\bigcup_{C\in Z} V(C)$, and rescale $\eps$ by a multiplicative factor of $3$ throughout the proof.
\end{clproof}

We are finally ready to build the requested bidecomposition. Let
$n = |V(G)|$ and fix $\eps = 1/\log{n}$. W.l.o.g. assume that
$\eps\le 1/8$, for otherwise $G$ is of constant size.  Define
weight functions on vertices of $G$ as follows: $\wei_1(u)=1$ for each
vertex $u$, and $\wei_2(u)=1$ for each $u\in S$ and $\wei_2(u)=0$ for
each $u\notin S$.  The bidecomposition is constructed using the
following recursive procedure which for $R\subseteq V(G)$ constructs a
bidecomposition of $G[R]$; we apply it initially to $R = V(G)$.

\begin{enumerate}[noitemsep,topsep=0pt]
\item If $R$ is empty, terminate and return an empty bidecomposition.
\item Otherwise, apply Claim~\ref{cl:balsep} to $H=G[R]$ with weight
  functions $\wei_1(\cdot)$ and $\wei_2(\cdot)$ restricted to
  $R$. This yields a partition $(A,X,B)$ of $R$. Apply the procedure
  recursively to $A$ and to $B$ in place of $R$, yielding
  bidecompositions of $G[A]$ and $G[B]$, respectively. Return a
  bidecomposition of $G[R]$ obtained by creating a root $r$, mapping
  all vertices of $X$ to $r$, and attaching the bidecompositions of
  $G[A]$ and $G[B]$ as children of $r$.
\end{enumerate}

Let $(T,\alpha)$ be the bidecomposition of $G$ obtained in this
manner.  As $\eps = 1/\log{n}$, property~\ref{p:parts} is clear from
the construction.  Therefore, we focus on proving
properties~\ref{p:depth} and~\ref{p:Sdepth}.

Consider any root-to-leaf path in $T$ and let
\[
  V(G) = R_0\supseteq R_1\supseteq R_2\supseteq \cdots\supseteq R_d
\]
be the consecutive sets $R$ considered by the procedure constructing
$(T,\alpha)$ along this path. By the construction we have
\[
  \card{R_i} \leq (1/2+\eps)^i\cdot n\qquad\textrm{for each
  }i\in\range{0}{d}.
\]
Since the procedure stops when $R$ becomes empty, we have that
$\card{R_d}\ge 1$, implying that
\[
  (1/2+\eps)^d\cdot n \geq 1
\]
or equivalently
\[
  d \leq - \log_{1/2+\eps}{n} = \frac{\log{n}}{ 1 +
    \log\pare{\frac{1}{1+2\eps}} }.
\]
Now observe that
\[
  \log\pare{\frac{1}{1+2\eps}} = \log \pare{ 1-\frac{2\eps}{1+2\eps}
  } \geq -\frac{4\eps}{1+2\eps} \geq - 4\eps.
\]
Here, the first inequality follows from $\log (1-x)\geq -2x$ for
$x\in [0,1/2]$, which in turn follows from the concavity of function
$t\mapsto \log t$.  Therefore, we conclude that
\[
  d \leq \frac{\log{n}}{1-4\eps} ~\le~ \log{n} \cdot (1+8\eps) =
  \log{n} + 8,
\]
where the second inequality follows from $1 \le (1-4\eps)(1+8\eps)$
being true for $\eps\le 1/8$. We conclude that the height of
$T$ is at most $\log{n} + 8$, so property~\ref{p:depth} is verified.

The proof of property~\ref{p:Sdepth} is analogous: instead of any
root-to-leaf path in $T$, we consider any path from the root to a
node $x$ satisfying $\alpha^{-1}(x) \cap S \neq \emptyset$. This
concludes the proof of the existential statement.

\medskip

We now discuss the algorithmic aspects. In the following, $k$ is considered a fixed constant.

It is easy to implement the procedure described in the proof of Claim~\ref{cl:balanced-trees} in linear time by processing the tree $T$ in a bottom-up manner: one keeps track of the total weight of unmarked descendants and when this count reaches $\eps\cdot \wei(T)$, the current node is added to $S$ and the count is zeroed. The proof of Claim~\ref{cl:epsseps} essentially boils down to applying Claim~\ref{cl:balanced-trees} to a tree decomposition of width $k$, with a weight function that can be computed in linear time directly from the definition. As for a fixed $k$, an optimum-width tree decomposition of a graph of treewidth at most $k$ can be found in linear time~\cite{Bod96a}, this gives a linear time algorithm for finding the set $Z$ promised by Claim~\ref{cl:epsseps}.

The procedure described in the proof of Claim~\ref{cl:balsep-abstract} can be easily
implemented in time linear in $|\Omega|$.
By combining this with the discussion of Claim~\ref{cl:epsseps} from the previous paragraph, for Claim~\ref{cl:balsep} we obtain an algorithm that runs in linear time.

Finally, the construction of the final bidecomposition amounts
to applying (the algorithm of) Claim~\ref{cl:balsep} recursively. It can be easily seen that graphs considered in the subcalls at every level of the recursion are disjoint, so the total amount of work used in those subcalls is linear in the size of the input graph. Since the recursion depth is bounded by $\log n$, we conclude that the algorithm runs in time $O(n\log n)$.
\end{proof}

\section{Case of a short path}
\label{sec:small-diam}

We now move to the first step of the proof of Theorem~\ref{thm:main}.
Hence, from now on we fix an efficiently flat class $\Cc$, and we let $w\in \N$ be the constant given by the efficient flatness of $\Cc$. In the following we treat $w$ as a constant, hence all the constants hidden in the $O(\cdot)$-notation may depend on $w$.

We now use our
understanding of schemes for graphs of bounded treewidth in order to lift it to graphs from $\Cc$
that can be embedded into $H\stimes P$, where $H$ has bounded treewidth and $P$ is a relatively short path. This intermediate result will be exploited in the next section in the general labeling scheme for $\Cc$.

\begin{lemma}\label{lem:main-diam}
  Graphs from $\Cc$ admit a
  labeling scheme of length $\log{n} + \log{d} + O(\log\log{n})$,
  where we assume that the Encoder is given a graph $G\in \Cc$ together with a subgraph embedding of $G$ into a graph $H\stimes P$, where $H$ has treewidth at most $w$ and $P$ is a path of length $d$. The Encoder runs in polynomial time and the Decoder in constant time.

  Moreover, if the graph $G$ is provided together with a vertex subset
  $Q$, then the Encoder may assign to the vertices of $Q$ labels of
  length at most $\log{|Q|} + \log{d} + O(\log\log{n})$.
\end{lemma}

\begin{proof}
  We first focus on proving the initial statement, without the
  additional vertex subset $Q$. At the end we shall argue how the
  refined statement can be obtained using property~\ref{p:saving} of Theorem~\ref{thm:stronger-tw}.

  Let $G\in \Cc$ be the input graph, where $G$ has $n$ vertices.
  We assume that we are also given a subgraph embedding $\phi$ of $G$ into $H\stimes P$, where $H$ has treewidth at most $w$ and $P$ has length $d$. As argued, by removing vertices not participating in the image of $G$ under $\varphi$, we may assume that $|V(H)|\leq n$ and $|V(P)|=d+1\leq n$.
  We identify the vertices of $P$ with numbers $\{0,1,\ldots,d\}$.
  
  Since $H$ has treewidth $w=O(1)$,
  we may apply Theorem~\ref{thm:stronger-tw} to $H$.
  Thus, in polynomial time we can compute a
  labeling $\kappa(\cdot)$ defined on vertices of $H$ with
  labels of length $\log{n} + O(\log\log{n})$, for which we have a Decoder working in constant time. 
  
  Now, we define a labeling $\lambda(\cdot)$ of $G$ as follows. 
  Take any $u\in V(G)$ and let $(v,i)=\phi(u)$, where $v\in V(H)$ and $i\in \{0,1,\ldots,d\}$.
  Then the label $\lambda(u)$ consists of:
  
\begin{itemize}[noitemsep,topsep=0pt]
    \item the label $\kappa(v)$; 
    \item the number $i$, written in binary;
    \item a $3(w+1)$-bit \emph{adjacency code}, which we define in a moment.
\end{itemize}

The first two pieces of information above are of variable length, so
we add to the label a prefix of (fixed) length $2\log\log{n}$ that
encodes their lengths, so that they can be extracted from the label in
constant time.
Clearly, the total length of any label constructed in this way is bounded by $\log n + \log d + O(\log \log n)$.

It remains to describe the adjacency code and how the decoding is
going to be performed based on it. Recall that, by
property~\ref{p:ids}, every vertex of $v$ of $H$ is assigned an
identifier $\iota(\kappa(v))$ so
that from $\kappa(v)$ one can compute a set $\Gamma(\kappa(v))$ of at
most~$w$ identifiers with the following property: $v$ and $v'$ are
adjacent in $H$ if and only if
$\iota(\kappa(v))\in \Gamma(\kappa(v'))$ or
$\iota(\kappa(v))\in \Gamma(\kappa(v'))$. By ordering identifiers
lexicographically, we may assume that sets returned by $\Gamma(\cdot)$
are organized as lists\footnote{In the original scheme of~\cite{GL07}, $\Gamma(\cdot)$ sets are organized into dictionary so that membership can be tested in constant time, independently of the size of $\Gamma(\cdot)$. This refinement does not matter here since the size is bounded by~$w=O(1)$.}.  Observe that two vertices $u$ and $u'$ of $G$
may be adjacent only if the following two assertions hold:
denoting $\phi(u)=(v,i)$ and $\phi(u')=(v',i')$, we must have that
\begin{itemize}[noitemsep,topsep=0pt]
\item $v$ and $v'$ are equal or adjacent in $H$; and
\item $|i-i'|\leq 1$.
\end{itemize}
Hence, the adjacency code assigned to a vertex $u$ of $G$ stores the following information:
denoting $(v,i)=\phi(u)$, for each
$v'\in \{v\}\cup \Gamma(\kappa(v))$ and $t\in \set{-1,0,1}$, we record whether $u$ is adjacent
to the unique vertex $u'$ with $\phi(u') = (v',i+t)$, provided it exists. Note that there is at most one $u'$ as above, because $\phi$ is injective.

Given the above description, the decoding can be performed as follows.
Suppose we are given labels $\lambda(u)$ and $\lambda(u')$ of two
vertices $u,u'\in V(G)$. Denoting $(v,i)=\phi(u)$ and $(v',i')=\phi(u')$, from these labels we may consecutively compute:
\begin{itemize}[noitemsep,topsep=0pt]
    \item numbers $i$ and $i'$;
    \item labels $\kappa(v)$ and $\kappa(v')$;
    \item lists $\Gamma(\kappa(v))$ and $\Gamma(\kappa(v'))$; and
    \item identifiers $\iota(\kappa(v))$ and $\iota(\kappa(v'))$.
\end{itemize}
Next, we check whether $\iota(\kappa(v))=\iota(\kappa(v'))$, or
$\iota(\kappa(v))\in \Gamma(\kappa(v'))$, or $\iota(\kappa(v'))\in \Gamma(\kappa(v))$. If
this is not the case, then $u$ and $u'$ are not adjacent in $G$, because $v$ and $v'$ neither equal nor adjacent in $H$.  Otherwise,
we check whether $i-i'\in \set{-1,0,1}$. Again, if
this is not the case, then $u$ and $u'$ are not adjacent in $G$, because $i$ and $i'$ are neither equal nor adjacent in $P$. Otherwise,
whether $u$ and $u'$ are adjacent can be read from the adjacency code
of $\lambda(u)$ or of $\lambda(u')$, depending on which identifier
belongs to which list.

From the above description it is clear that the Encoder for this
labeling scheme runs in polynomial time, while the Decoder runs in
constant time.  This concludes the proof of the initial statement,
without the additional vertex subset $Q$. For the additional
statement, we simply apply the following modification:  we use property~\ref{p:saving} of Theorem~\ref{thm:stronger-tw} to ensure
that in the labeling $\kappa(\cdot)$, the vertices of $H$ that appear on the first coordinates of $\phi(Q)$
receive labels of length $\log{|Q|} + O(\log\log{n})$. Thus, in
$\lambda(\cdot)$ the vertices of $Q$ receive labels of total length at
most $\log{|Q|} + \log{d} + O(\log\log{n})$.
\end{proof}

\begin{remark}\label{rem:saving}
  In the labeling scheme of Lemma~\ref{lem:main-diam}, we reserve $\ceil{\log(d+1)}$ bits in
  the label of each vertex $u$ to store the {\em{$P$-coordinate}} of $\varphi(u)$, that is, the number $i$ where $\varphi(u)=(v,i)$. Observe that we may modify the scheme so that for
  vertices the $P$-coordinate $i$ is either $0$ or $d$, this
  piece of information takes $O(1)$ bits. Namely, using $3$ first bits
  we store $i$ is equal to $0$, $1$, $d-1$, $d$, or lies between
  $2$ and $d-2$. Then the value of $i$ is recorded using
  $\ceil{\log(d+1)}$ additional bits only when it is between $1$ and
  $d-1$. It is easy to see that using this way of storing the $P$-coordinates, the Decoder can verify
  whether two given $P$-coordinates (decoded from the labels) differ by at most $1$ (and then what is their difference), even when the
  Decoder does not know the value of $d$ in advance.  We will use this
  optimization in the next section.
\end{remark}

\begin{remark}
  The proof of flatness of planar graphs given by Dujmovi\'c{} et al.~\cite{DEJWW19} actually provides a subgraph embedding of every planar graph $G$ into a graph of the form $H\stimes P$, where $H$ is a graph of treewidth at most $8$ and $P$ is a path whose length is bounded by the diameter of $G$. Hence, from Lemma~\ref{lem:main-diam} we can infer that planar graphs admit a labeling scheme of length $\log n+\log d+O(\log \log n)$, where $d$ is the diameter of the input graph.
\end{remark}

\section{General case}
\label{sec:planar}

Finally, we are ready to prove our main result, Theorem~\ref{thm:main}.

\begin{proof}[of Theorem~\ref{thm:main}]
  Let $G = (V,E)\in \Cc$ be the input graph on $n$ vertices.
  Let
  $$d = \ceil{n^{1/3}}.$$
  W.l.o.g. we assume that $d\ge 3$ (or $n>8$). 
  
  By efficient flatness of $\Cc$, we may compute in polynomial time a graph $H$ of treewidth at most $w$, a path $P$, and a subgraph embedding $\varphi$ of $G$ into $H\stimes P$. As before, by removing unnecessary vertices we may assume that $|V(H)|\leq n$ and $|V(P)|\leq n$. Letting $\ell$ be the length of $P$, we again identify the vertices of $P$ with numbers $\{0,1,\ldots,\ell\}$. 
  
  For $i\in \{0,1,\ldots,\ell\}$ let 
  $$L_i= \varphi^{-1}(\{(v,i)\colon v\in V(H)\}),$$
  and for $a\in \range{0}{d-1}$ let
  $$W_a = \bigcup_{i\in \N\colon i\equiv a\bmod d} L_i.$$
  Note that $\{L_0,L_1,\ldots,L_\ell\}$ and $\{W_0,W_1,\ldots,W_{d-1}\}$ are partitions of $V$.
  
  When speaking about sets $W_a$, we consider indices modulo $d$.
  Then one of the sets $W_a\cup W_{a+1}$ is small in the following
  sense.

  \begin{claim}\label{cl:light-layer}
    There exists $a\in \range{0}{d-1}$ such that
    $|W_a\cup W_{a+1}| \le 2n^{2/3}$.
  \end{claim}
  \begin{clproof}
    Observe that $\sum_{a\in [0,d)} |W_a\cup W_{a+1}| = 2n$ because
    every vertex belongs to exactly two of the sets $W_a\cup W_{a+1}$. Hence, for
    some $a\in \range{0}{d-1}$ we have
    $|W_a\cup W_{a+1}| \le 2n/d \le 2n^{2/3}$.
  \end{clproof}

  Partition the edges of $G$ into $E_1$ and $E_2$ as follows:
  \begin{itemize}[nosep]
      \item $E_1$ comprises all edges with one endpoint in $W_a$ and the other in $W_{a+1}$;
      \item $E_2$ comprises all the remaining edges.
  \end{itemize}
  Next, define subgraphs $G_1$ and $G_2$ of $G$ as follows:
  $$
    G_1 = (W_a\cup W_{a+1},E_1) \qquad\textrm{and}\qquad G_2 =
    (V,E_2).$$
  We first show that $G_1$ is a very simple and small graph.
  
  \begin{claim}\label{cl:G1}
    The graph $G_1$ has at most $2n^{2/3}$ vertices and treewidth at
    most~$2w+1$.
  \end{claim}
  \begin{clproof}
    The bound on the number of vertices of $G_1$ is directly implied
    by Claim~\ref{cl:light-layer}. 
    
    For the treewidth bound, note
    that every connected component of $G_1$ is a subgraph of the graph
    $G[L_i\cup L_{i+1}]$ for some $i\in \N$.  
    Next, observe that mapping $\varphi$ restricted to $L_i\cup L_{i+1}$ witnesses that $G[L_i\cup L_{i+1}]$
    is a subgraph of $H\stimes K_2$, where $K_2$ is the graph consisting of two adjacent vertices.
    It is easy to see that if $H$ has treewidth at most $w$, then $H\stimes K_2$ has treewidth at most $2w+1$: in a tree decomposition of $H$ of width at most $w$ just replace every vertex $v$ with the two copies of $v$ in $H\stimes K_2$. Since treewidth does not grow under taking subgraphs, we conclude that $G[L_i\cup L_{i+1}]$ has treewidth at most $2w+1$. Hence every connected component of $G_1$ has treewidth at most $2w+1$, so we can conclude the same about the whole $G_1$.
  \end{clproof}

  We now analyze the graph $G_2$. The idea is to apply Lemma~\ref{lem:main-diam}, so we need to find a subgraph embedding of $G_2$ into a graph $H'\stimes P'$, where $H'$ has bounded treewidth while $P'$ is a {\em{short}} path. We do it as follows.
  
  \begin{claim}\label{cl:G2}
    In polynomial time one can compute a graph $H'$ of treewidth at most $w$ and a subgraph embedding $\varphi'$ of $G_2$ into $H'\stimes P'$, where $P'$ is a path of length $d-1$. Moreover, we have $$W_a=\varphi'^{-1}(\{(v,d-1)\colon v\in V(H')\})\qquad\textrm{and}\qquad W_{a+1}=\varphi'^{-1}(\{(v,0)\colon v\in V(H')\}).$$
  \end{claim}
  \begin{clproof}
  Let $H'$ be the graph obtained by taking $n$ disjoint copies of $H$; clearly the treewidth of $H'$ is at most $w$. We assume that every vertex of $H'$ is represented as a pair $(v,j)$, where $v\in V(H)$ and $j\in \range{0}{n-1}$ is the index of the copy of $H$ in $H'$. Let $P'$ be the path of length $d-1$, whose vertices are indexed with numbers $0,1,\ldots,d-1$. Consider the following mapping $\varphi'$ from $V(G)$ to $V(H'\stimes P')$:
  for $u\in V(G)$, denoting $(v,i)=\phi(u)$ and $j=i+d-a-1$, we set
  $$\varphi'(u)= (\ (v\,,\,j\div d)\, ,\, j\bmod d\ ),$$
  where $j\div d = \lfloor j/d\rfloor$. Note that the assertion that $W_a=\varphi'^{-1}(\{(v,d-1)\colon v\in V(H')\})$ and $W_{a+1}=\varphi'^{-1}(\{(v,0)\colon v\in V(H')\})$ follows directly from the definition.
  
  It is straightforward to verify that $\varphi'$ is a subgraph embedding from $G_2$ to $H'\stimes P'$: the subgraph $G_2[L_0\cup \ldots \cup L_a]$ is mapped to the first copy of $H$ (times $P'$), the subgraph $G_2[L_{a+1}\cup \ldots \cup L_{a+d}]$ is mapped to the second copy of $H$ (times $P'$), and so on. Note here that in $G_2$ there are no edges between $L_a$ and $L_{a+1}$, nor between $L_{d+a}$ and $L_{d+a+1}$, and so on.
  \end{clproof}
  
We can now use Claims~\ref{cl:G1} and~\ref{cl:G2} to give the promised
labeling scheme.  First, by Claim~\ref{cl:G1} and
Theorem~\ref{thm:bnd-tw}, for the graph $G_1$ we may compute a
labeling $\lambda_1$ with labels of length at most
$\frac{2}{3}\log{n}+O(\log\log{n})$. Second, by Claim~\ref{cl:G2} and
Lemma~\ref{lem:main-diam}, for $G_2$ we may compute\footnote{A careful reader might be worried at this point that the graph $H'$ produced by Claim~\ref{cl:G2} may have as many as $\Omega(n^2)$ vertices. However, since $G$ has at most $n$ vertices, we may again remove vertices of $H'$ that do not participate in the image of $G$ under $\varphi'$, thus bringing $|V(H')|$ to at most $n$. In fact, this is what happens at the beginning of the proof of Lemma~\ref{lem:main-diam}.} a labeling
$\lambda_2$ with labels of length at most
$\log{n}+\log d+O(\log\log{n})=\frac{4}{3}\log{n}+O(\log\log{n})$. 
Moreover, we may construct this labeling so that all
vertices of $W_a\cup W_{a+1}$ receive shorter labels, namely of length
at most $\frac{2}{3}\log{n}+O(\log\log{n})$. Here, we use
Remark~\ref{rem:saving} together with the second statement of Claim~\ref{cl:G2} in order to reduce the $\log d$ summand to $O(1)$, and we use $Q=W_a\cup W_{a+1}$ as the prescribed set of at
most $2n^{2/3}$ vertices in order to reduce the $\log{n}$ summand
to $\frac{2}{3}\log{n}+O(1)$.

Now, for any vertex $u$ of $G$, we define its label $\lambda(u)$ as follows:
\begin{itemize}
\item If $u\notin W_a\cup W_{a+1}$, then $\lambda(u)=\lambda_2(u)$.
\item If $u\in W_a\cup W_{a+1}$, then $\lambda(u)$ is the
  concatenation of $\lambda_1(u)$ and $\lambda_2(u)$.
\end{itemize}
In the second case, in order to be able to decode $\lambda_1(u)$ and
$\lambda_2(u)$ from $\lambda(u)$, we append $\log\log{n}$ bits that
indicate the length of $\lambda_1(u)$. Also, we add one bit indicating
the case into which the vertex $u$ falls.

Thus, in the first case $\lambda(u)$ is of length
$\frac{4}{3}\log{n}+O(\log\log{n})$, while in the second it is of
length
\[
  \frac{2}{3}\log{n} +O(\log\log{n})+\frac{2}{3}\log{n} +O(\log\log
  n)+\log\log{n}=\frac{4}{3}\log{n}+O(\log\log{n}) ~.
\]
Hence, the length of the labeling scheme is as claimed.
For the implementation of the Decoder, given labels of two vertices
$u$ and $u'$, we simply read labels of $u$ and $u'$ in $\lambda_2$ and
$\lambda_1$ (if applicable) and check whether they are adjacent either
in $G_1$ or in $G_2$. This concludes the construction of the labeling
scheme.

The above construction can be directly translated to an implementation
of the Encoder in polynomial time and the Decoder in constant time. In
case of Claim~\ref{cl:light-layer}, note that an index $a$ satisfying
the claim can be found in polynomial time by checking all the integers
between $0$ and $d-1$ one by one.
\end{proof}

\begin{remark}
Note that the statement of Theorem~\ref{thm:main} considers the class $\Cc$ fixed, hence the factors hidden in the $O(\cdot)$ notation depend on the constant $w$ given by the efficient flatness of $\Cc$. It is not hard to verify that the obtained dependence on $w$ is linear, that is, the length of the labeling scheme provided by Theorem~\ref{thm:main} is $\frac{4}{3}\log n + c\cdot w\cdot \log \log n$ for some absolute constant $c\in \N$.
\end{remark}

\section{Conclusions}\label{sec:conc}

We gave an upper bound of $(\frac{4}{3}+o(1))\log{n}$ for
  the length of labeling schemes for any efficiently flat class of graphs.
  This result applies to the class of planar graphs --- which were our main motivation --- but also encompasses more general classes such as graphs embeddable in a fixed surface, or $k$-planar graphs for any fixed $k$. Subsequent to our work, Dujmovi\'c et al.~\cite{DEJGMM20} announced a construction of labeling schemes for any efficiently flat class of graphs with the optimum length $(1+o(1))\log{n}$. Their approach also relies on the structure theorem of~\cite{DJMPx19}, but is much more involved than the one presented here. Also, it leads to a non-constant-time implementation of the Decoder.
  
  So far, the extent of these results is delimited by the flatness of the considered classes of graphs. As discussed in~\cite{DJMPx19}, this property is unfortunately not enjoyed by every proper minor-closed graph class, but holds for every class of nearly embeddable graphs without apices (formally, $(0,g,k,p)$-nearly embeddable graphs for fixed $g,k,p\in \N$). Such graphs are the basic building blocks in the Structure Theorem for proper minor-closed classes of Robertson and 
  Seymour~\cite{RS04b}. This gives hope for extending the existence of labeling schemes of length $(1+o(1))\log{n}$ to all proper minor-closed classes through the Structure Theorem. In fact, such a line of reasoning was successfully applied in~\cite{DJMPx19} to obtain upper bounds on the queue number in proper minor-closed classes. In the case of labeling schemes, combining the labelings along a tree decomposition into nearly-embeddable parts seems to be the main issue.
  
\paragraph{Acknowledgements.}
We are grateful to Vida Dujmovi\'c{} for pointing out that our original proof, which was tailored to the cases of planar graphs and of graphs embeddable in a fixed surface, can be also performed on the level of generality of arbitrary flat classes of graphs. Apart from generalizing the results, this greatly clarified and streamlined the presentation of the reasoning.

A part of this research was completed at the 7th Annual Workshop on
Geometry and Graphs held at Bellairs Research Institute in March 2019.

\bibliographystyle{my_alpha_doi}
\bibliography{soda}

\end{document}